\documentclass[conference]{IEEEtran}
\IEEEoverridecommandlockouts
\usepackage{cite}
\usepackage{amsmath,amssymb,amsfonts}
\usepackage{algorithmic}
\usepackage{graphicx}
\usepackage{textcomp}
\usepackage{xcolor}
\def\BibTeX{{\rm B\kern-.05em{\sc i\kern-.025em b}\kern-.08em
    T\kern-.1667em\lower.7ex\hbox{E}\kern-.125emX}}
\usepackage{makecell}
    
\begin{document}

\title{
Virtual Sensing for Solder Layer Degradation and  Temperature Monitoring in IGBT Modules
\thanks{
This study was conducted within the PowerizeD project, funded by the KDT Joint Undertaking under grant 101096387 (with support from Horizon Europe and National Authorities).}
}

\author{
Andrea Urgolo\thanks{$^*$Corresponding authors; these authors contributed equally.}$^{*}$, Monika Stipsitz$^{*}$, Hèlios Sanchis-Alepuz\\
\textit{Embedded Systems, Silicon Austria Labs GmbH, Graz, Austria}\\
\{andrea.urgolo, monika.stipsitz, helios.sanchis-alepuz\}@silicon-austria.com}

%--------------- TODO ! ----------------------------
% \usepackage{siunitx}
\newcommand{\todo}[1]{\textcolor{red}{TODO: #1}}
%--------------- new ----------------------------
\newcommand{\lowr}[1]{_\textrm{#1}}
\newcommand{\degree}{^{\circ}}
% ----------------------------------------------------

\maketitle

\begin{table}[b] % full-width float at the bottom
% \centering
\footnotesize
% Preprint submitted to \textit{2025 International Conference on System Reliability and Safety.}
\copyright~2025 IEEE. Personal use of this material is permitted. Permission from IEEE must be obtained for all other uses, in any current or future media, including reprinting/republishing this material for advertising or promotional purposes, creating new collective works, for resale or redistribution to servers or lists, or reuse of any copyrighted component of this work in other works. Published in: 2025 9th International Conference on System Reliability and Safety (ICSRS). DOI: 10.1109/ICSRS68021.2025.11422051
\end{table}

\begin{abstract}
Monitoring the degradation state of Insulated Gate Bipolar Transistor (IGBT) modules is essential for ensuring the reliability and longevity of power electronic systems, especially in safety-critical and high-performance applications. However, direct measurement of key degradation indicators -- such as junction temperature, solder fatigue or delamination -- remains challenging due to the physical inaccessibility of internal components and the harsh environment. In this context, machine learning-based virtual sensing offers a promising alternative by bridging the gap from feasible sensor placement to the relevant but inaccessible locations. This paper explores the feasibility of estimating the degradation state of solder layers, and the corresponding full temperature maps based on a limited number of physical sensors. Based on synthetic data of a specific degradation mode, we obtain a high accuracy in the estimation of the degraded solder area (1.17 \% mean absolute error), and are able to reproduce the surface temperature of the IGBT 
with a maximum relative error of 4.56\% (corresponding to an average relative error of 0.37\%). 
\end{abstract}


\section{Introduction}

Power electronic systems play a crucial role in a wide range of industrial and automotive applications, where efficiency, reliability, and long operational lifetimes are key requirements. Among the most critical components in these systems are Insulated Gate Bipolar Transistor (IGBT) modules, which are widely used for high-power switching due to their favorable electrical characteristics. However, like all semiconductor devices operating under substantial thermal loads, IGBT modules are susceptible to degradation mechanisms, particularly in the solder layers that connect the silicon die to the substrate and the substrate to the baseplate.

Accurate monitoring of temperature and degradation state within IGBT modules is essential to ensure safe operation and to implement predictive maintenance strategies. Yet, direct measurement of internal physical variables, such as junction temperature or the extent of solder layer fatigue and delamination, is infeasible in practice. This is due to the harsh operating environment, space constraints, and the fact that the most critical regions for degradation are physically inaccessible to conventional sensors.
To overcome this, various degradation monitoring approaches have been proposed (see \cite{Huang2021} for an overview): Model-based methods estimate remaining useful life using accessible measurements and statistical device data, while data-driven methods adapt to specific devices using runtime sensor histories. However, these approaches typically lack the spatial resolution necessary to track localized degradation phenomena, such as partial solder delamination or non-uniform heat spreading.

In this context, machine learning-based virtual sensing has emerged as a promising alternative. By learning the relationships between a limited number of physical sensor readings and the underlying spatial temperature or degradation states, virtual sensors have the potential to infer inaccessible quantities without the need for intrusive instrumentation \cite{Zhao2021}. 
In the literature, virtual sensors have been widely adopted in a variety of fields and are typically categorized according to the modeling paradigm: first-principle (physics-based), black-box (data-driven), and grey-box (hybrid) approaches~\cite{brunello2021virtual,urgolo2024virtual}.

In this paper, we investigate to what extent ML-based virtual sensing methods can be used to accurately reconstruct internal solder degradation states and temperature fields in IGBT modules, using only a limited number of externally accessible sensor measurements. To answer this question, we generate synthetic thermal data from degraded IGBT modules and train ML models to infer both the degraded solder area and the resulting temperature distributions for two typical degradation modes. In contrast to previous simulation-based studies, we generate a training dataset containing hundreds of different degradation states to cover the variability of actual systems. Our results demonstrate high predictive accuracy, suggesting a viable path forward for real-time, non-intrusive health monitoring of power electronic systems.

\section{Methods}

This section describes the machine learning and dataset generation methods used in this work.

\subsection{Machine Learning Models}\label{sec:ml-models}

The core objective of this work is to leverage machine learning to reconstruct internal solder degradation states and temperature distributions in IGBT modules from a limited set of physically accessible temperature measurements. To this end, we consider and compare a range of supervised regression models, from established baselines to more advanced neural architectures, for both solder area fraction prediction and maximum chip temperature estimation.

As baselines, we include standard linear regression, random forest regression, and XGBoost regression models. These approaches offer interpretable predictions and serve as widely accepted references in the field. Linear regression provides a straightforward benchmark for linear correlations between sensor readings and target variables, while random forest and XGBoost—based on ensembles of decision trees—are capable of capturing more complex, nonlinear relationships and interactions among input features.

Building upon these baselines, we adopt a feed-forward neural network (FFNN) as the main data-driven model for both prediction tasks. The FFNN takes as input the set of temperature readings from selected sensor locations or sensor grids and outputs either the estimated solder area fraction or the predicted maximum chip temperature, depending on the task. The architecture consists of multiple fully connected layers with non-linear activations and dropout regularization. For tasks involving spatially structured input, such as predicting temperature fields from grid or graph-structured sensor data, we further employ graph neural network (GNN) models—specifically, attention-based graph architectures—which leverage node and edge features to capture spatial relationships on the chip surface.

All models are implemented using open-source Python libraries, including \texttt{scikit-learn} for the baselines~\cite{scikit-learn}, \texttt{XGBoost} for gradient-boosted regression~\cite{chen2016xgboost}, \texttt{PyTorch} for the neural networks~\cite{paszke2019pytorch} and \texttt{PyTorch Geometric} for the graph-based models~\cite{fey2019fast}.

While the models discussed above are purely data-driven and rely exclusively on observed sensor measurements for inference, further improvements in accuracy and robustness can be achieved by incorporating physical domain knowledge into the training process. To this end, we extend the standard loss function with a physics-based regularization term derived from the steady-state heat equation, effectively transitioning from a purely black-box approach to a hybrid (grey-box) modeling framework, 
by penalizing discrepancies between the predicted and FE-computed integral heat fluxes.
We have previously observed that adding physics information as a regularizer to the training loss improves the generalization capabilities of NNs in predicting temperature distributions \cite{Sabathiel2024,Stipsitz2022a}. For this work, we develop such a regularizing term for the 2D graph structures described in Sec.~\ref{sec:data-extraction}. The details of the derivation can be found in Appendix~\ref{appendix_heateq}.

\subsection{Dataset generation}\label{sec:dataset}

The developed virtual sensing methods are applied to a generic IGBT system consisting of one chip and one diode. A dataset was generated using steady-state Finite Element (FE) simulations. The system was simplified as a layered structure with dimensions and material properties approximated from literature and datasheets \cite{Hu2019,An2023,Pedersen2012,Infineon2024} (see Tabs.~\ref{Tab:dimensions} and \ref{Tab:materials}). Boundary conditions adapted from \cite{Wang2022} were applied, consisting of Robin boundary conditions representing forced heat extraction at the bottom of the base plate (heat transfer coefficient $\alpha=200 \cdot 10^3 \mathrm{W}/(\mathrm{K}\mathrm{m}^2$)) and on all other outer surfaces ($\alpha=12.5 \mathrm{W}/(\mathrm{K}\mathrm{m}^2)$). In both cases using the randomized external temperature $T\lowr{ext}$ per system. Power loss was represented as a volumetric heat source (corresponding to heat source power $h$) applied to the chip die.

\begin{figure}[htbp]
\centering
\begin{minipage}{0.2\textwidth}
    \centering
    \includegraphics[width=\textwidth]{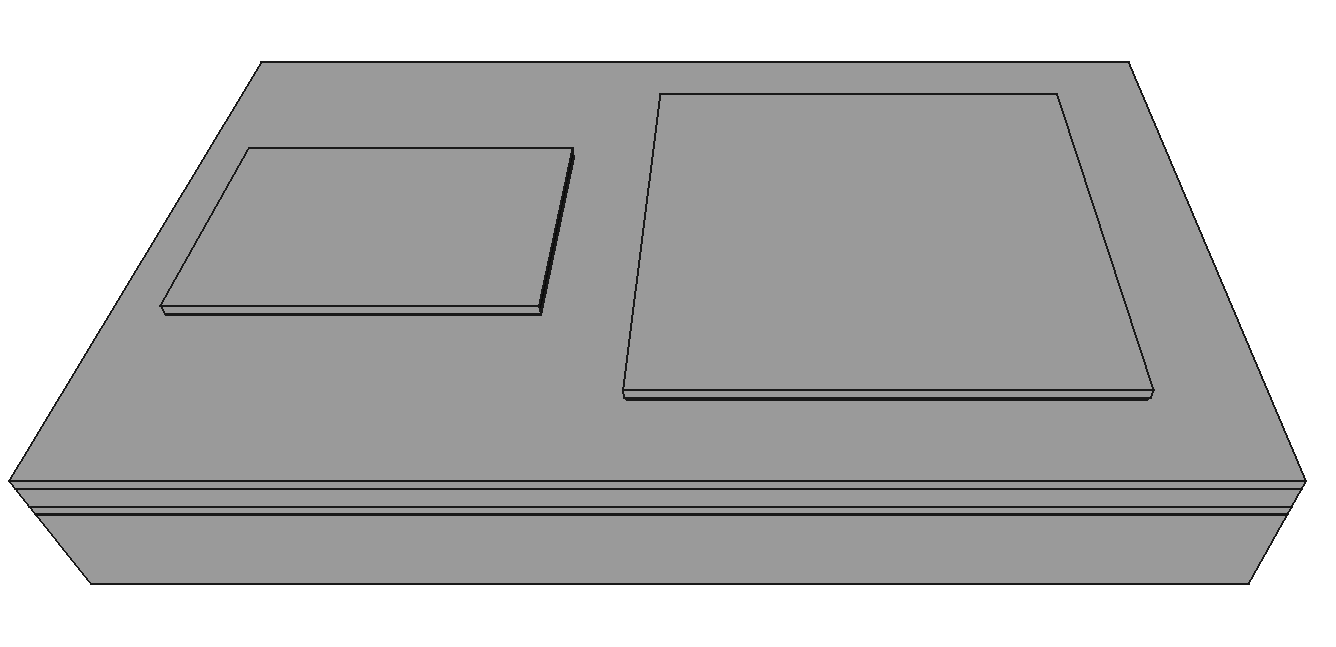} \\
    (a) System geometry
\end{minipage}%
\begin{minipage}{0.15\textwidth}
    \centering
    \includegraphics[width=\textwidth]{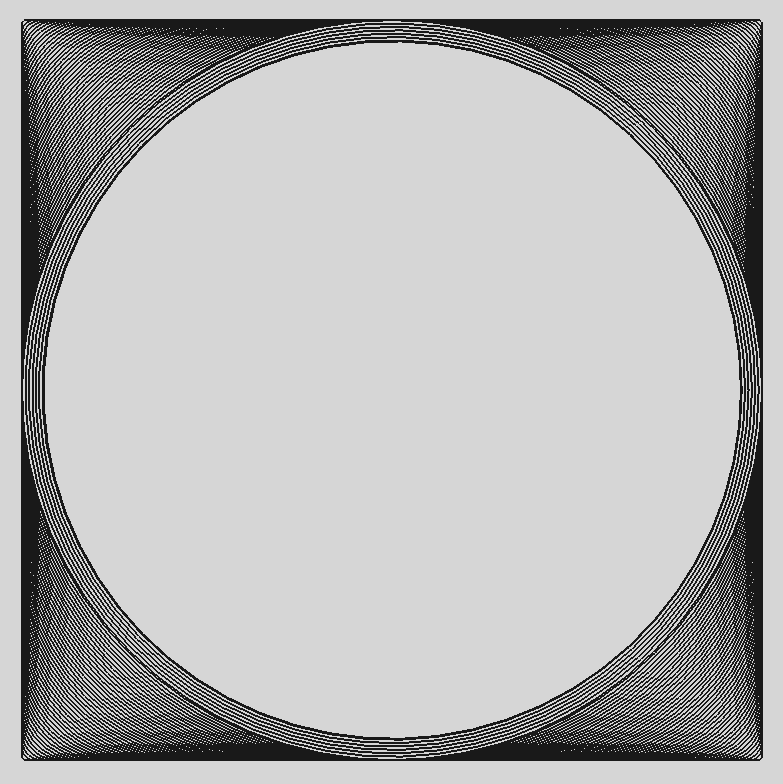} \\
    (b) Delamination
\end{minipage}%
\begin{minipage}{0.15\textwidth}
    \centering
    \includegraphics[width=\textwidth]{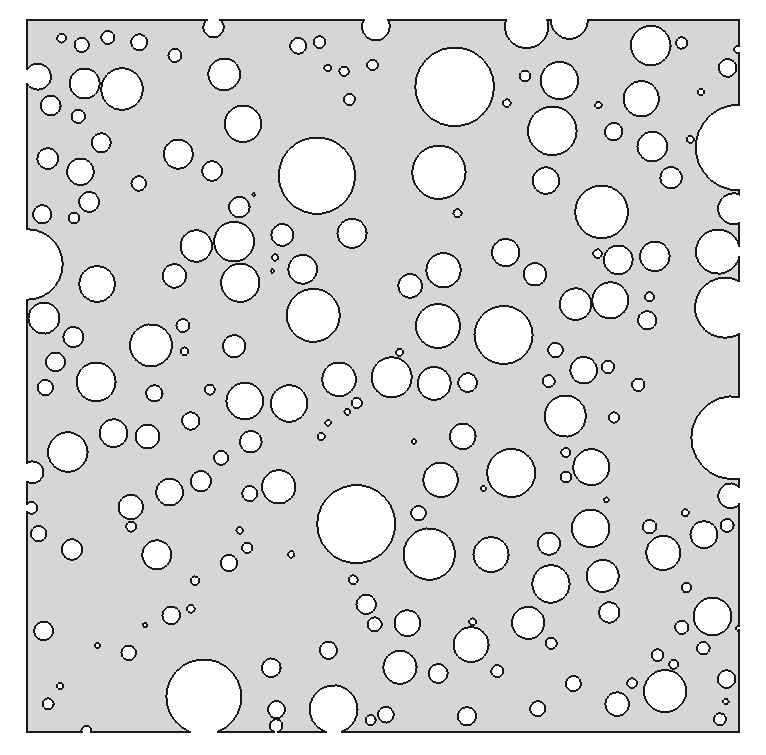} \\
    (c) Voiding
\end{minipage}\\
\caption{Idealized geometries used in this study: (a) Overview of the system with the diode on the left and the chip on the right. (b) Delamination scenario: View from top onto the chip solder layer (other components of the system are hidden) at different degradation states. The different states are shown on top of each other. (c) Solder voiding scenario: Exemplary void configuration at a lost solder layer fraction of $\alpha=25.6\%$ (view from top onto the chip solder layer).}
\label{fig:geometry}
\end{figure}

In this work, we focus on illustrating the potential of virtual sensing for two of the main degradation scenarios of the chip solder layer \cite{Huang2019}, which are reproduced by varying the geometry of the solder layer before meshing: \textbf{(1) Delamination:} In short, this degradation mode is characterized by the delamination of the quadratic solder layers, starting at the corners and finally leading to circular cross-sections of the remaining solder layers. We mimic this failure scenario in the FE simulations by idealized geometries at different stages of degradation of the solder layers (similar to \cite{Hu2019}). In total 55 different geometries were generated (see Fig.~\ref{fig:geometry}b) -- from undegraded solder layer to approximately 30\% loss of solder layer area. \textbf{(2) Solder voiding:} In this scenario, the solder layer degrades uniformly by an increasing number of voids which are forming as the degradation increases. We use a simple model, motivated by previous works \cite{Du2020,Li2024,Alavi2024,Le2016}, assuming that the voids are cylindrical in shape and going over the full depth of the solder layer. First, we select a goal degradation state which is given by a total fraction of lost solder layer. Then, we sample void sizes from a lognormal distribution taken from \cite{Le2016}, Tab.~6 and place them at random locations -- which are not overlapping with any previously defined voids -- in the solder layer, see Fig.~\ref{fig:geometry}c for an example void configuration. In the FEM simulation air was assigned as material to the voids.

In both scenarios, a training dataset was generated leveraging an automated workflow: based on a randomly selected solder layer fraction (sampled from a uniform distribution with the ranges given in Tab.~\ref{Tab:random_properties}) a corresponding CAD geometry and conformal FE mesh was constructed using the scriptable open-source tools \texttt{FreeCAD} and \texttt{gmsh}. Then, for each system a random $h$ and $T\lowr{ext}$ were selected based on a uniform distribution within the ranges in Tab.~\ref{Tab:random_properties}. The FEM simulations were conducted using the open-source FEM solver \texttt{ElmerFEM}. In total, the training datasets for the delamination and the voiding degradation scenarios consisted of approximately 700 and 1000 individual FEM simulations, respectively. Additionally, two evaluation datasets were generated with fixed parameters ($h = 400 W$, $T\lowr{ext}=20 \degree C$) mimicking increasing states of degradation of a single system. For scenario (1) this means that all geometries shown in Fig.~\ref{fig:geometry}b are simulated, while for scenario (2) the dataset starts with a random configuration with total degraded solder layer fraction $\alpha_0 = 1\%$, then randomly selected voids grow with a growth rate of $20\%$ and new voids appear such that $\alpha$ grows by increments of $1\%$ until finally $\alpha = 30\%$ is reached.

\begin{table}[tbp]
\caption{Dimensions of the simplified IGBT system.}
% \begin{center}
\centering
\begin{tabular}{|c|c|c|c|c|}
\hline
& \multicolumn{3}{|c|}{\textbf{Dimensions}} & {\textbf{Material}} \\
% \cline{2-5}
\hline
 Component & \textbf{\textit{Length[mm]}}& \textbf{\textit{Width[mm]}}& \textbf{\textit{Height[mm]}} & \\
\hline
 Chip & 13 & 13 & 0.3  & Si \\
 Diode & 10 & 7 & 0.3  & Si \\
\hline
 Chip solder & 13 & 13 & 0.05 & SnAgCu \\
 Diode solder & 10 & 7 & 0.05 & SnAgCu \\
\hline
 DCB Copper & 30 & 18 & 0.3 & Cu \\
 DCB Ceramic & 30 & 18 & 0.7 &  $\mathrm{Al}_{2}\mathrm{O}_3$ \\
 DCB Copper & 30 & 18 & 0.3 & Cu \\
\hline
 DCB Solder & 30 & 18 & 0.05 & SnAgCu \\
\hline
 Baseplate & 30 & 18 & 3 & Cu \\
\hline
\end{tabular}
\label{Tab:dimensions}
% \end{center}
\end{table}

\begin{table}[tbp]
\caption{Material properties used in the FEM simulations.}
% \begin{center}
\centering
\begin{tabular}{|c|c|c|c|}
\hline
% & \multicolumn{3}{|c|}{\textbf{Dimensions}} & {\textbf{Material}} \\
% \cline{2-5}
% \hline
 Property & \textbf{\textit{Density $\rho$}}& 
 \textbf{\textit{Heat conductivity $k$}} &
  \textbf{\textit{Specific heat capacity $c\lowr{p}$}} \\
 % units
 Unit  & $\mathrm{kg}/\mathrm{m}^3$& 
  $\mathrm{W}/(\mathrm{K}\mathrm{m})$ &
$\mathrm{J}/(\mathrm{kg} \mathrm{K})$ \\
\hline
% Air & 1.2 & 0.024 & 1004.9 \\
% \hline
 Si & 2330 & 100 & 741  \\
\hline
 SnAgCu & 7400 & 78 & 0.22 \\
 \hline
 Cu & 8930 & 380 & 381 \\
\hline
 $\mathrm{Al}_{2}\mathrm{O}_3$ & 3987 & 25 & 850 \\
\hline
\end{tabular}
\label{Tab:materials}
% \end{center}
\end{table}

\begin{table}[htbp]
\caption{Variability of simulations: For each simulation of the dataset each property is independently sampled from a uniform distribution.}
% \begin{center}
\centering
\begin{tabular}{|c|c|c|}
\hline
Property & Minimum & Maximum \\
\hline
% & \multicolumn{3}{|c|}{\textbf{Dimensions}} & {\textbf{Material}} \\
% \cline{2-5}
% \hline
 Removed solder layer fraction $a$ (\%) & 0 & 30 \\
 \hline
 Heat source power $h$ (W) & 100 & 700  \\
\hline
 External temperature $T\lowr{ext}$ ($\degree$C) & 15 & 40 \\
\hline
\end{tabular}
\label{Tab:random_properties}
% \end{center}
\end{table}

\subsection{Data Extraction}\label{sec:data-extraction}

For each simulation in the dataset, a graph-based representation of the chip surface is constructed to enable spatially resolved machine learning models. The extraction process consists of three main steps: ($i$) identification of the three physical reference sensor locations, ($ii$) systematic selection of surface nodes, and ($iii$) construction of the graph topology and feature matrices.

\paragraph*{($i$) Reference Sensor Locations}
Three reference temperature points, denoted as $T_1$, $T_2$, and $T_3$, are defined based on physical accessibility and relevance to monitoring strategies:
\begin{itemize}
    \item $\mathbf{T_1}$ is positioned at the geometric center of the chip surface, $(x_1, y_1, z_1) = (20.5\,\mathrm{mm},\,9.0\,\mathrm{mm},\,4.7\,\mathrm{mm})$, corresponding to the nominal junction temperature ($T_\mathrm{j}$), which can be indirectly estimated in real devices via temperature-sensitive electrical parameters (TSEP).
    \item $\mathbf{T_2}$ is placed at the same $(x, y)$ coordinates as $T_1$ but at the bottom of the module, $(x_2, y_2, z_2) = (20.5\,\mathrm{mm},\,9.0\,\mathrm{mm},\,0.0\,\mathrm{mm})$, representing a temperature that could be measured at the baseplate for calibration or redundancy.
    \item $\mathbf{T_3}$ is located near the chip periphery on the top copper layer of the direct copper bonded (DCB) substrate, at $(x_3, y_3, z_3) = (30.0\,\mathrm{mm},\,9.0\,\mathrm{mm},\,4.35\,\mathrm{mm})$, reflecting a feasible position for a thin-film or foil-type sensor accessible in practice.
\end{itemize}
These coordinates are mapped onto the simulation mesh by identifying, for each reference point, the mesh node closest to the target coordinates, using a minimal Euclidean distance criterion. This approach ensures that physical sensor locations are robustly and consistently represented across all simulated device instances.

\paragraph*{($ii$) Surface Node Extraction}
To construct the graph, nodes are selected from the FE mesh corresponding to the chip’s top surface. Candidate nodes are filtered according to three geometric constraints:
\begin{itemize}
    \item \textbf{Height:} 
    The $z$-coordinate must match the top surface ($z \approx 4.7\,\mathrm{mm}$), within a specified tolerance ($\pm 10^{-5}\,\mathrm{m}$, i.e., $\pm 0.01\,\mathrm{mm}$).
    \item \textbf{Lateral Extent:} Only nodes within a predefined rectangular region covering the active chip area are included. Specifically, nodes must satisfy $x \in [14.0,\,27.1]\,\mathrm{mm}$ and $y \in [2.5,\,15.5]\,\mathrm{mm}$.
    \item \textbf{Spatial Separation:} To avoid redundant or clustered node placement, a greedy selection procedure is used whereby each new node is included only if it is at least $ L = 0.5\,\mathrm{mm}$ from all previously selected nodes. This minimum separation $L$ defines the characteristic node spacing on the chip surface.
\end{itemize}
This results in a set of target surface nodes, augmented by the three reference points (if not already included), yielding the complete set of graph nodes. Each node is assigned a feature vector comprising local quantities (e.g., temperature, position, material properties) as available from the FE simulation.

\paragraph*{($iii$) Graph Topology and Edge Construction}
Edges between nodes are constructed to encode spatial relationships and facilitate message passing in downstream graph neural network models. The procedure is as follows:
\begin{itemize}
    \item \textbf{Surface Node Edges:} Each surface node is connected to all other surface nodes within a maximum Euclidean distance of $0.866\,\mathrm{mm}$ (i.e., $L\sqrt{3}$ with $L = 0.5\,\mathrm{mm}$, as defined above), as determined by a k-d tree nearest-neighbor search. Edges are undirected and encode local spatial connectivity.    
    \item \textbf{Reference–Surface Connections:} To ensure the integration of information from the reference sensors, bidirectional edges are added between each reference node and all surface nodes.
    \item \textbf{Edge Features:} For each edge, a feature vector comprising the relative spatial displacement $\Delta \mathbf{r} = \mathbf{r}_j - \mathbf{r}_i$, where $\mathbf{r}_i$ and $\mathbf{r}_j$ denote the coordinates of the connected nodes $i$ and $j$, is computed and stored. This enables the network to leverage geometric relationships during data preprocessing and training.

\end{itemize}
The resulting graph, therefore, contains a set of nodes corresponding to selected physical locations on the chip surface, including reference sensors, and a set of spatially meaningful edges suitable for graph-based machine learning. Node and edge features are extracted directly from the FE solution fields for each simulation instance.

The extraction and processing pipeline was implemented in \texttt{Python} exploiting the \texttt{meshio} library~\cite{schloemer_meshio} to read the finite element results from the unstructured mesh data stored in VTU format.
To construct the spatial neighborhood graph, efficient nearest-neighbor searches were performed using \texttt{scipy.spatial.KDTree}. The resulting graphs, including node and edge features, were represented as instances of \texttt{torch\_geometric.data.Data} from the \texttt{PyTorch Geometric} library~\cite{fey2019fast} for subsequent use in graph-based regression and prediction tasks.

\begin{figure*}[tb]
    \centering
    \includegraphics[width=.78\textwidth]{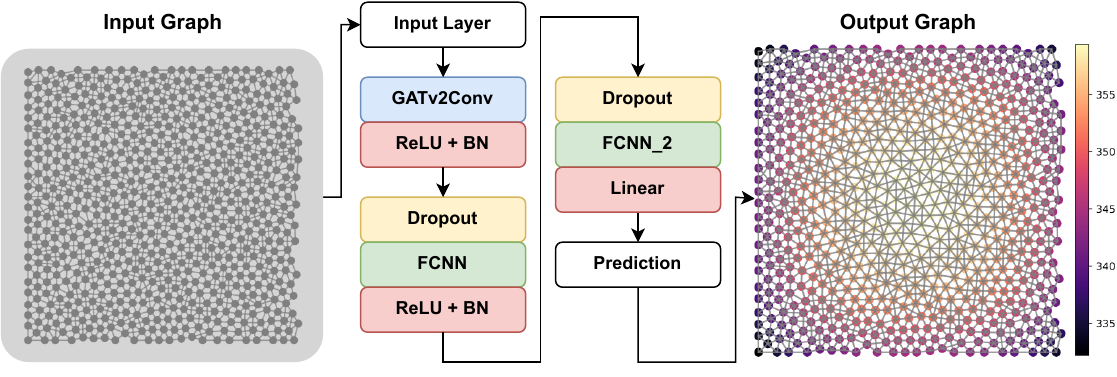}
    \caption{
        Architecture of the edge-conditioned graph attention network for chip surface temperature prediction. An example of a graph extracted from the dataset, highlighting the temperature distribution (in Kelvin) across chip surface nodes, is shown on the right as the output of the predictive model.}
    \label{fig:gn_arch}
\end{figure*}

\section{Results}

This section reports the results obtained for solder area fraction prediction and temperature monitoring, evaluating 
the performance of the proposed models
across all experimental settings for both delamination and solder voiding scenarios.

\subsection{Solder Layer Fraction Prediction}
\label{sec:solder_fraction_prediction}

The prediction of the solder layer fraction is a key step in assessing the degradation state of the IGBT modules in our virtual sensing framework. We focus on the delamination dataset, as described in Sec.~\ref{sec:dataset}, which consists of a series of FE-based simulations representing progressive shrinkage of the solder area from the initial undegraded state down to approximately 70\% of the original area. For each synthetic chip instance, we construct a single feature vector composed of the temperatures of three reference sensors, denoted as $T_1, T_2,$and $ T_3$  respectively (as defined in Sec.~\ref{sec:data-extraction}). The first temperature $T_1$ is related to the junction temperature $T_\mathrm{j}$. In real applications, this value can be estimated from electrical operating conditions using established methods such as temperature-sensitive electrical parameters (TSEP). In our experiments, we optionally add controlled Gaussian noise of magnitude $\gamma_{T_\mathrm{j}}$ to $T_1$ in order to emulate the estimation uncertainty on the junction temperature.

In addition to the three temperature readings, the heat source power $h$ applied to the chip die is included as an input feature. This value represents the total dissipated power during device operation and directly affects the overall temperature profile of the chip and solder layer. 

The final input vector for each chip is thus $\mathbf{x} = [T_1, T_2, T_3, h]$. The prediction target is the solder layer area fraction $\alpha \in [0, 1]$, computed as the ratio between the current and initial solder areas.

Prior to training, we aggregate the data into a matrix of features $\mathbf{X} \in \mathbb{R}^{N \times 3}$ and a target vector $\mathbf{y} \in \mathbb{R}^N$, where $N$ is the total number of simulated samples. Feature values are rescaled to $[0,1]$ using MinMax normalization, $x_{i,k}^\mathrm{(scaled)} = (x_{i,k} - \min(x_k)) / (\max(x_k) - \min(x_k) + \epsilon)$, where $\epsilon$ is a small constant for stability. The dataset is split into training and validation subsets, using a 90\%/10\% partitioning with random shuffling, ensuring a uniform and unbiased selection of samples across the degradation range.

For the regression task, we compared several predictive models. The primary model is a FFNN with multiple hidden layers and ReLU activations, each followed by batch normalization and dropout for regularization.

The architecture is parameterized by the number of layers, units per layer, dropout rate, learning rate, and $L_2$ weight decay. Specifically, we performed a grid search  on the validation split over the following discrete sets of hyperparameter values: number of layers $\in \{1, 2, 3\}$, units per layer $\in \{64, 128, 256, 512\}$, dropout rate $\in \{0.0, 0.1, 0.3, 0.5\}$, learning rate $\in \{10^{-5}, 10^{-4}, 10^{-3}\}$, $L_2$ weight decay $\in \{10^{-6}, 10^{-5}, 10^{-4}, 10^{-3}, 10^{-2}\}$, and batch size $\in \{8, 12, 16, 32, 64\}$. Training is performed using the Adam optimizer and mean squared error (MSE) loss. In order to promote convergence and avoid stagnation during training, a learning rate scheduler based on validation loss plateaus is employed, with a patience of 10 epochs and decay factor $0.9$.

All experiments are run with a set of $10$ fixed random seeds for reproducibility. Based on the results of the grid search, the final neural model used for solder fraction prediction consists of a FFNN with an input layer of $3$ features, $2$ hidden layers with $512$ and $256$ units, respectively, and a single output unit. Each hidden layer includes batch normalization, ReLU activation, and dropout regularization with a rate of $0.3$. The model is optimized using Adam with a learning rate of $10^{-4}$ and $L_2$ weight decay of $10^{-3}$. The batch size was set to 16, and training was performed for up to 1000 epochs.

\begin{figure}[htbp]
\centering
\includegraphics[width=\linewidth]{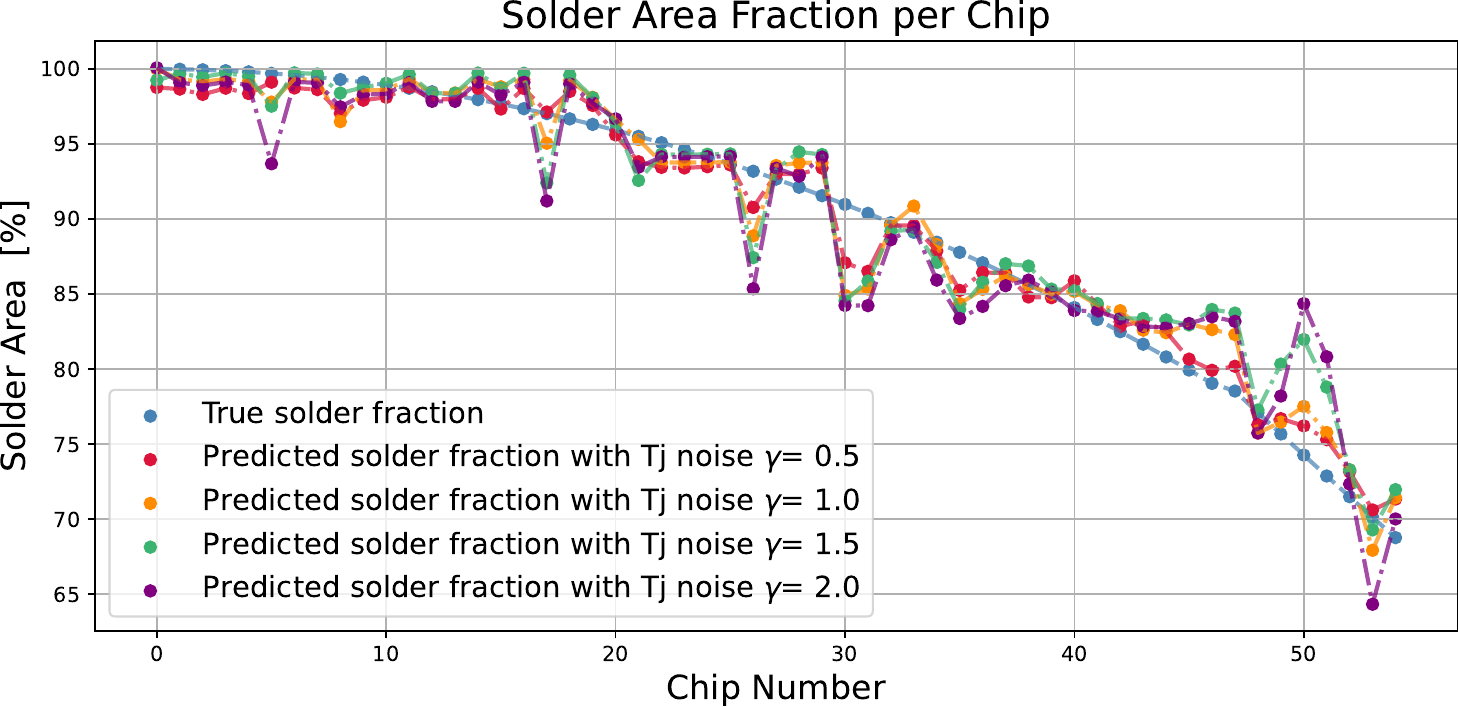} 
\caption{Predicted and true solder area fraction per chip in the delamination test dataset for increasing levels of synthetic $T_\mathrm{j}$ noise maignitude ($\gamma$).}
\label{fig:solder_area_prediction}
\end{figure}

Fig.~\ref{fig:solder_area_prediction} compares the FFNN-predicted solder area fraction to the true values for all chips in the delamination evaluation dataset, across different levels of additive Gaussian noise applied to the $T_\mathrm{j}$ input ($\gamma = 0.5$, $1.0$, $1.5$, and $2.0$). As the input noise increases, the model's accuracy degrades gradually. At $\gamma=0.5$, the mean absolute error (MAE) is $1.17\%$ and the maximum error is below $3.89\%$, indicating highly precise estimation. For $\gamma=2.0$, MAE rises to $2.13\%$ and the maximum error reaches $10.09\%$. These results should be interpreted in light of the typical uncertainty encountered when estimating $T_\mathrm{j}$ in power modules. In practice, the error in junction temperature estimation is commonly on the order of $\pm 1$~K,~\cite{Wang2022}. Notably, applying additive Gaussian noise with standard deviation $\gamma=0.5$ means that, according to the properties of the normal distribution, 95\% of perturbations fall within $\pm 1.0$~K of the true value. This range closely matches the real-world measurement uncertainty for $T_\mathrm{j}$, making $\gamma=0.5$ a realistic and representative setting. Therefore, in the remainder of this work, we use $\gamma=0.5$ as the default noise magnitude for all experiments involving junction temperature perturbations.

% Results with different GAMMA
% Predicted solder fraction for Tj $\gamma$ =  0.5 MAE: 0.011743313632905483, max error 0.03888469934463501
% Predicted solder fraction for Tj $\gamma$ =  1.0 MAE: 0.015084932558238506, max error 0.06053406000137329
% Predicted solder fraction for Tj $\gamma$ =  1.5 MAE: 0.018837345764040947, max error 0.07697248458862305
% Predicted solder fraction for Tj $\gamma$ =  2.0 MAE: 0.021287640556693077, max error 0.1008564829826355

As baselines, to evaluate the effectiveness of the FFNN approach, we compared its performance on the test set against the scikit-learn implementations of standard linear regression (ordinary least squares) and random forest regression~\cite{pedregosa2011scikit}, as well as an XGBoost regressor model~\cite{chen2016xgboost}. 
All of them were trained on the same input features and evaluated using identical splits and preprocessing, with default parameters used for computational convenience and to allow a rapid comparison. 
Tab.~\ref{tab:solder_fraction_regressors} summarizes the results. 
The FFNN consistently outperforms the other approaches, achieving a markedly lower maximum absolute error and thus demonstrating greater robustness in this evaluation scenario.

\begin{table}[htbp]
\caption{Test set performance of different regression models for solder area fraction prediction.}
% \begin{center}
\centering
\begin{tabular}{|c|r|r|r|r|r|}
\hline
\textbf{Model} & \makecell{\textbf{MAE} \\ \text{[\%]}} & \makecell{\textbf{Max Error} \\ \text{[\%]}} & \makecell{\textbf{Mean Rel.} \\ \textbf{L1 Error}} & \makecell{\textbf{Max Rel.} \\ \textbf{L1 Error}} \\
\hline
FFNN & \textbf{1.1743} & \textbf{ 3.8885 } & \textbf{0.0134} & \textbf{0.0427}  \\
\hline
Linear Reg. & 3.3556 & 9.0335 &  0.0382 & 0.1074 \\
\hline
Random Forest & 3.6106 & 25.9245 & 0.0434 & 0.3557  \\
\hline
XGBoost Reg. & 3.7606 & 23.1912 & 0.0443 & 0.3182 \\
\hline
\end{tabular}
\label{tab:solder_fraction_regressors}
% \end{center}
\end{table}

% FFNN MSE 0.00021177523012738675 MAE 0.011743313632905483 max error 0.03888469934463501 Rel L1 error 0.013371305540204048 Max Rel L1 error 0.04274268448352814
% Linear Regression MSE 0.0016090936 MAE 0.033556193113327026 max error 0.090334951877594 Rel L1 error 0.038173843175172806 Max Rel L1 error 0.10742389410734177
% Random Forest MSE 0.005240429778743907 MAE 0.036105636564168056 max error 0.25924499988555905 Rel L1 error 0.043441246012062666 Max Rel L1 error 0.3557074849954884
% XGBoost Regressor MSE 0.004239512 MAE 0.037605736404657364 max error 0.23191207647323608 Rel L1 error 0.04428219422698021 Max Rel L1 error 0.31820425391197205

\subsection{Maximum Chip Temperature Prediction}\label{sec:max_chip_temp_prediction}

This section addresses the prediction of the maximum chip temperature in the IGBT module under progressive solder layer delamination, leveraging graph-based neural networks. The task consists in estimating the maximum temperature within the active chip region, based on spatially distributed measurements and features encoded in node and edge attributes of the corresponding FEM-derived graph.

This experiment is performed on the delamination scenario introduced in Sec.~\ref{sec:dataset}, using the same train/validation split as for the solder fraction regression. 
For each chip instance, a graph is constructed with node features encoding local temperature (for reference nodes only, masked as $-1$ elsewhere), spatial coordinates, distance from chip center, and global features such as the predicted solder fraction and maximum heat source $h$. Edges are assigned 
attributes comprising distance and unit direction between source and destination nodes. As in the earlier experiments, all features are min-max scaled using statistics computed over the training set.

The main predictive model is a graph neural network based on edge-conditioned graph attention layers (GAT), as illustrated in Fig.~\ref{fig:gn_arch}. The architecture consists of a single \texttt{GATv2Conv} layer (as implemented in~\cite{fey2019fast}), which processes node and edge features, followed by a two-layer FFNN head comprising batch normalization, dropout, and a linear layer for per-node temperature prediction. 

We also investigate the effect of incorporating a physics-based loss term, enforcing agreement between the predicted and FEM-computed integral heat fluxes according to the discretized steady-state heat equation, as outlined in~\ref{appendix_heateq}.

Hyperparameter tuning is conducted via grid search on the validation split. Specifically, the following discrete sets are considered: number of GAT layers $\in \{1, 2, 3\}$, hidden units per head $\in \{32, 64, 128\}$, number of attention heads $\in \{4, 8, 12, 16\}$, dropout rate $\in \{0.0, 0.2, 0.4\}$, batch size $\in \{8, 12, 16, 32\}$, learning rate $\in \{10^{-4}, 5 \times 10^{-5}, 10^{-5}\}$, and $L_2$ weight decay $\in \{10^{-5}, 10^{-4}, 10^{-3}\}$. The output FFNN head is parameterized by one or two hidden layers (width $\in \{32, 64, 128\}$) and uses batch normalization and dropout. For models including the heat equation term, the regularization weight is chosen from $\{10^{-4}, 10^{-5}, 10^{-6}, 10^{-7},10^{-8}\}$.

All models are trained with the Adam optimizer, using mean squared error as the primary loss, optionally augmented with the physics-based loss as described above. To improve convergence and mitigate overfitting, a learning rate scheduler is employed, reducing the learning rate by a factor of $0.5$ if the validation loss does not improve for $10$ epochs. The maximum number of epochs is set to $1000$. Training and evaluation are repeated with $10$ fixed random seeds for reproducibility.

After grid search, the optimal model configuration for the delamination dataset consists of a single GAT layer with $64$ hidden units and $12$ attention heads, and FFNN head with two hidden layers ($64$ units each). Dropout rates of $0.2$ (attention) and $0.4$ (MLP), learning rate $10^{-4}$, $L_2$ weight decay $10^{-3}$, batch size $12$, and heat equation regularization weight $10^{-5}$ yield the best validation results.

The effectiveness of this model has been quantitatively assessed and is reported in Tab.~\ref{tab:temp_regressors}, which compares its performance to several standard regression baselines.
Among all considered approaches, the GAT-based methods consistently achieve lower maximum errors than classical regression baselines, with the inclusion of the heat equation regularization further reducing both the average and worst-case errors. Specifically, the physics-regularized GAT model yields the lowest MAE ($1.44$~K) and maximum error ($20.88$~K), indicating improved reliability over both standard GAT and tree-based ensemble models.

\begin{table}[htbp]
\caption{Test set performance of different regression models for maximum chip temperature prediction.}
% \begin{center}
\centering
\begin{tabular}{|l|r|r|r|r|}
\hline
\textbf{Model} 
& \makecell{\textbf{MAE} \\ \textbf{[K]}}
& \makecell{\textbf{Max Error} \\ \textbf{[K]}}
& \makecell{\textbf{Mean Rel.} \\ \textbf{L1 Error}}
& \makecell{\textbf{Max Rel.} \\ \textbf{L1 Error}} \\
\hline
Linear Reg. & 14.9565 & 349.9039 & 0.0359 & 0.4145 \\
\hline
Random Forest & 2.3501 & 56.3082 & 0.0060 & 0.1245 \\
\hline
XGBoost Reg. & 4.0211 & 97.1093 & 0.0101 & 0.1501 \\
\hline
GAT & 4.4857 & 27.3205 & 0.0118 & 0.0668 \\
\hline
GAT + Heat Eq. & \textbf{1.4439} & \textbf{20.8752} & \textbf{0.0037} & \textbf{0.0456} \\
\hline
\end{tabular}
\label{tab:temp_regressors}
% \end{center}
\end{table}

\begin{figure}[htbp]
\hfill
\includegraphics[width=.98\linewidth]{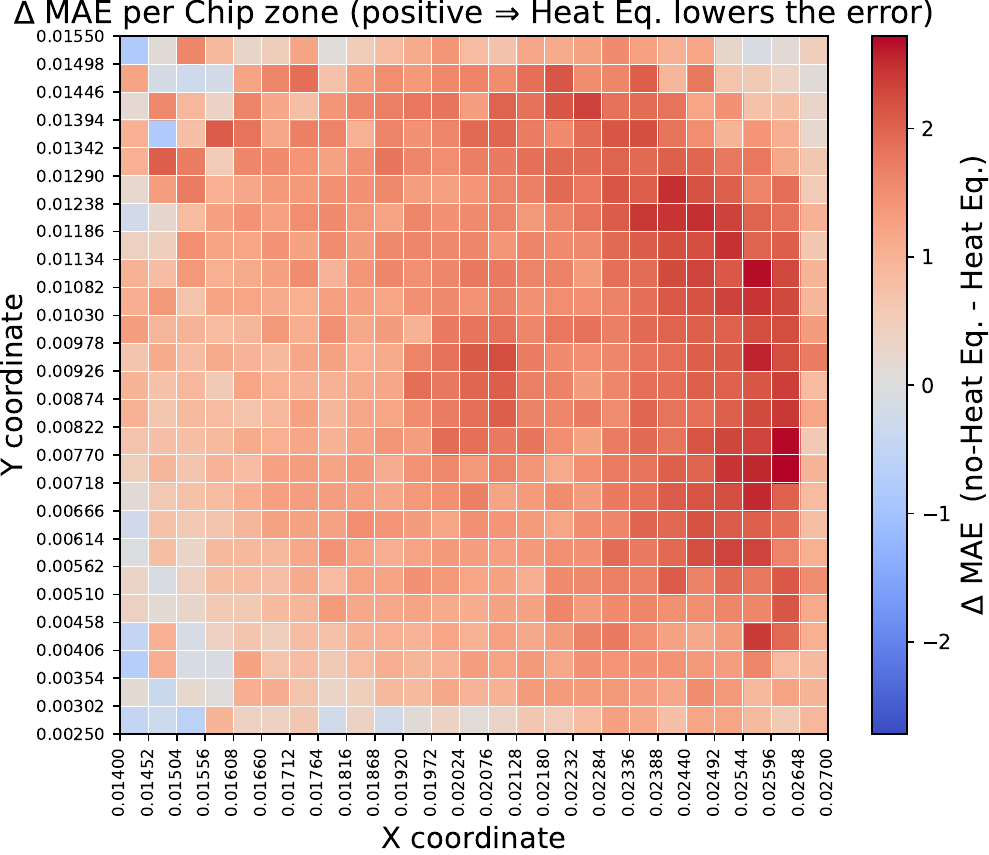} 
\caption{Difference in mean absolute error per chip zone between the standard graph network model
and the physics-regularized variant with the heat equation constraint, in the delamination test dataset. Errors are aggregated over a $25 \times 25$ spatial grid.}
\label{fig:mae_diff_per_zone}
\end{figure}

\begin{figure}[ht]
    \centering
    \includegraphics[width=0.48\textwidth]{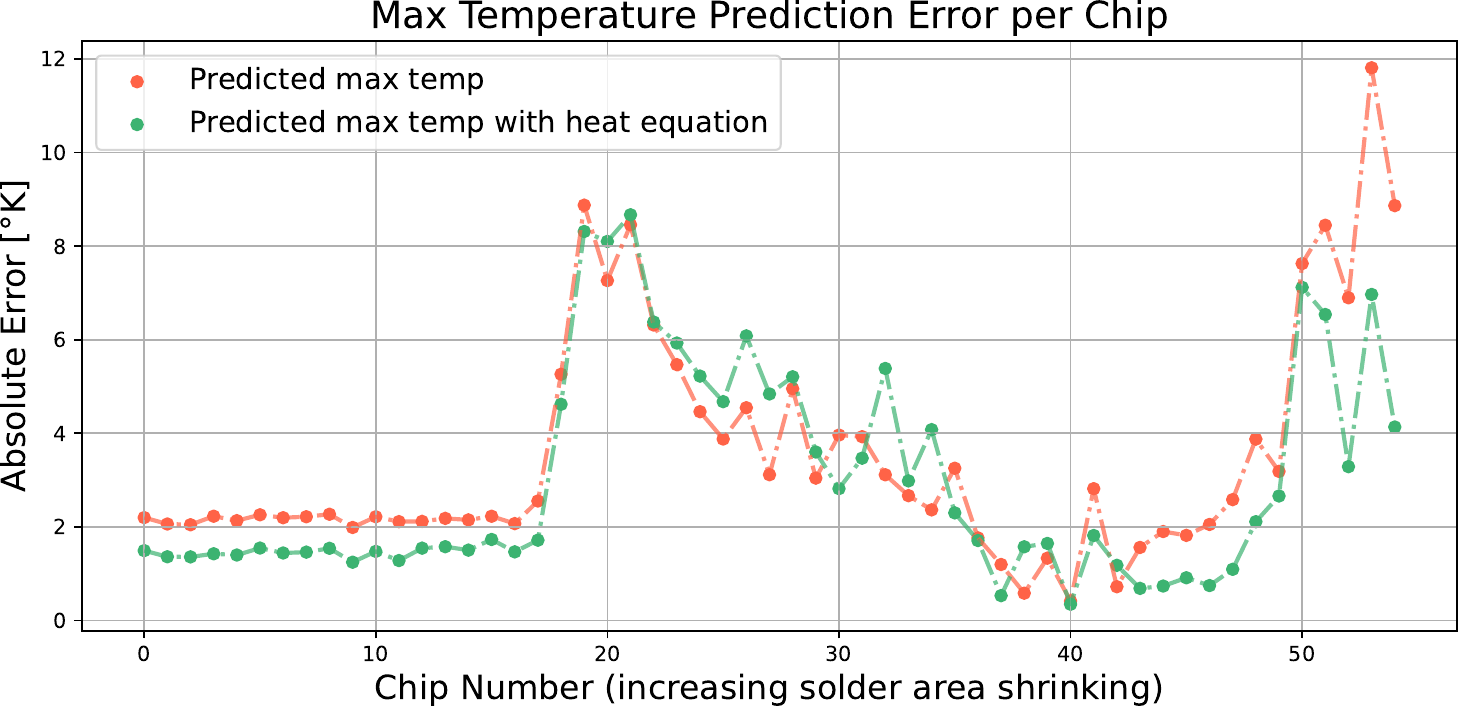}
    \caption{
        Absolute prediction errors for maximum chip temperature across the test set, comparing the standard graph network model and the physics-regularized variant with the heat equation constraint.
    }
    \label{fig:max_temp_per_chip}
\end{figure}

The effect of physics-based regularization can be further analyzed by examining the spatial distribution of prediction errors and the per-chip performance. Fig.~\ref{fig:mae_diff_per_zone} presents the difference in MAE per chip zone between the standard GAT and the variant regularized with the heat equation. 
To obtain this comparison, the chip surface is partitioned into a $25 \times 25$ grid of equally sized zones, within which the MAE is averaged over all predicted nodes located in each zone.
The regularized model achieves consistently lower errors across the vast majority of the chip surface, as indicated by the predominantly positive color scale. Only a few peripheral regions display negligible or slightly negative differences, suggesting rare cases where the constraint does not yield an improvement.

A complementary view is provided by the per-chip error curves in Fig.~\ref{fig:max_temp_per_chip}. The absolute prediction errors for maximum chip temperature are generally reduced for the physics-regularized GAT compared to the unconstrained baseline, with the largest benefits observed for the most degraded chips (i.e., those with the highest chip number and the smallest remaining solder area). Notably, the introduction of the heat equation constraint not only decreases the typical error, but also mitigates outlier errors, leading to a more stable prediction profile across all degradation states. Taken together, these results demonstrate that the incorporation of physical priors via the heat equation yields more accurate and robust temperature predictions, especially in challenging scenarios involving severe solder degradation.

\subsection{Solder Layer Fraction Prediction under Solder Voiding}
\label{sec:solder_fraction_prediction_voiding}

The prediction of the solder layer fraction 
becomes considerably more difficult in the case of solder voiding, as described in Sec.~\ref{sec:dataset}. In this degradation scenario, the solder layer is progressively perforated by randomly located voids, which lead to a spatially heterogeneous thermal field and significantly reduce the observability of global degradation from a limited number of surface temperature measurements.

Unlike the delamination case, using only the standard three reference sensor locations (as described in Sec.~\ref{sec:solder_fraction_prediction}) proved insufficient for accurate prediction: all tested machine learning models failed to learn a meaningful relationship between the sensor readings and the true solder area fraction. This  is consistent with the physical intuition that, in the presence of distributed voids, local temperatures become much less correlated with the total degraded area, especially when the sensors are placed far from the main sites of degradation.

To investigate the sensing requirements for this more complex scenario, we performed a simulation-based analysis of the effect of increasing the number and distribution of reference sensors. Specifically, we evaluated virtual sensor grids of varying density, all placed at the same height as the top copper layer of the DCB substrate ($z = 4.35$~mm), which is the uppermost metallization layer where temperature sensors (e.g., thin-film or foil resistive sensors) can realistically be integrated without major process modifications. Lower layers, such as the solder–baseplate interface ($z = 3.0$~mm) or the baseplate bottom ($z = 0$~mm), were also considered; however, preliminary tests and physical considerations indicated that measurements at these depths provide even less sensitivity to localized solder voids due to increased thermal averaging and reduced proximity to the critical solder layer.

For each configuration, the best-performing model from Sec.~\ref{sec:solder_fraction_prediction} (a FFNN with two hidden layers) was trained using input features corresponding to the temperatures measured at the vertices of the sensor grid, ranging from a single point 
% (1$\times$1), 
to $2\times2$, $3\times3$, $4\times4$, and $5\times5$ regular grids, all covering the active chip region. The same feature normalization and preprocessing steps were applied as in the delamination scenario.

\begin{table}[tbp]
\caption{Test set performance for solder area fraction prediction under voiding, as a function of sensor grid size.}
% \begin{center}
\centering
\begin{tabular}{|l|c|c|c|c|}
\hline
\textbf{Sensor grid} 
& \makecell{\textbf{MAE} \\ \textbf{[\%]}} 
& \makecell{\textbf{Max Error} \\ \textbf{[\%]}}
& \makecell{\textbf{Mean Rel.} \\ \textbf{L1 Error}}
& \makecell{\textbf{Max Rel.} \\ \textbf{L1 Error}} \\
\hline
$1 \times 1$  & 6.7893 & 15.1013 & 0.0877 & 0.2324 \\
$2 \times 2$  & \textbf{3.6961} & 10.9252 &  \textbf{0.0443} & \textbf{0.1104} \\
$3 \times 3$  & 3.7873 & \textbf{9.4469}  & 0.0490 & 0.1454 \\
$4 \times 4$  & 6.7480 & 19.0081 & 0.0856 & 0.2187 \\
$5 \times 5$  & 5.9869 & 15.0688 & 0.0804 & 0.2316 \\
\hline
$3$ reference  & 7.7235 & 17.9378 & 0.1009 & 0.2760 \\
\hline
\end{tabular}
\label{tab:voiding_solder_pred_results}
% \end{center}
\end{table}

Tab.~\ref{tab:voiding_solder_pred_results} summarizes the predictive accuracy achieved for each sensor configuration. As expected, a single reference sensor yields a high mean absolute error (MAE $\approx 6.8\%$) and maximum error exceeding $15\%$, confirming that the voiding problem is poorly observable from minimal sensor information. Introducing a $2\times2$ or $3\times3$ sensor grid reduces both MAE and maximum error by about half. 
Larger grids ($4\times4$, $5\times5$) do not lead to further improvements and, in some cases, introduce overfitting or noise amplification. This effect is likely due to the increased model capacity allowing the network to fit the specific patterns and sensor combinations in the training set, which do not necessarily generalize to the distinct test set conditions.
For reference, the final row of the table shows that using only the input from the three standard reference points, as in the delamination case, results in the worst predictive performance. This outcome highlights the need for greater sensor coverage to effectively address the loss of observability caused by spatially distributed voids.

\begin{figure}[t]
\centering
\includegraphics[width=\linewidth]{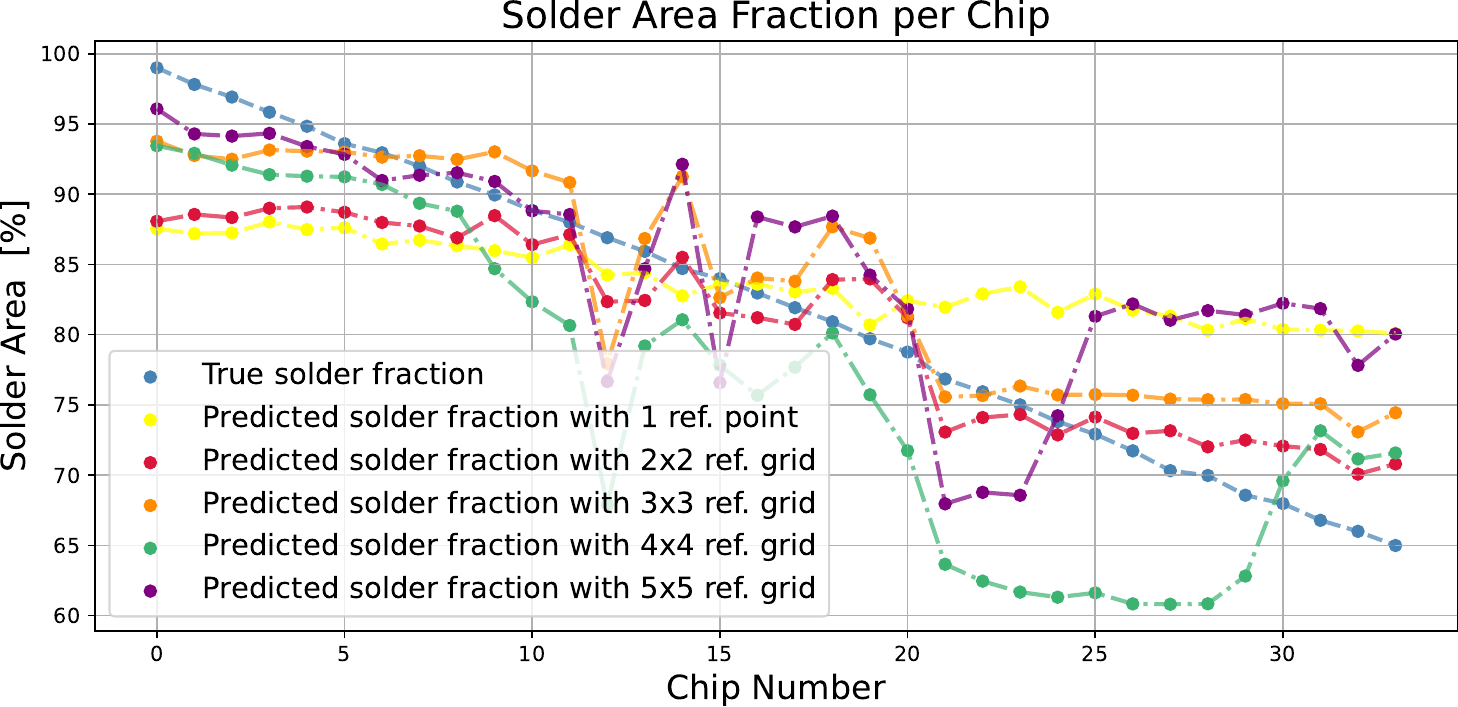}
\caption{
Predicted and true solder area fraction per chip for different sensor grid configurations in the solder voiding scenario test set.
}
\label{fig:solder_area_voids_grid}
\end{figure}

Fig.~\ref{fig:solder_area_voids_grid} shows the predicted solder area fraction per chip for each configuration, compared to the true degradation trajectory. The results clearly illustrate the sharp performance transition as the sensor grid density increases: only when a moderate spatial coverage is achieved does the model become capable of tracking the degradation trend.

\subsection{Maximum Chip Temperature Prediction under Solder Voiding}

In the solder voiding scenario, accurately estimating the maximum chip temperature poses unique challenges. The presence of multiple, spatially distributed voids disrupts the thermal field and reduces the effectiveness of surface measurements, making it substantially harder to infer peak temperatures from accessible sensor data. While the use of GNNs proved highly effective in the delamination setting, their application to the voiding dataset is hindered by both practical and methodological factors. Specifically, the extreme heterogeneity and non-local nature of the voiding-induced thermal field, 
resulted in unstable GNN training and poor generalization in preliminary experiments. Furthermore, the increased model capacity of graph networks appeared to exacerbate overfitting when only sparse or moderate sensor grids were available, as the network could memorize specific local patterns in the training set without learning robust global relationships.

Consequently, for this scenario, we adopt an approach based on a FFNN architecture, mirroring the setup described in Sec.~\ref{sec:solder_fraction_prediction_voiding}. The FFNN was trained to predict the maximum chip temperature using as input the full set of grid sensor readings (i.e., temperature values at the $N$ grid points), augmented with the predicted solder area fraction from the previous section as an additional feature. The same normalization and preprocessing 
as in the solder fraction task was applied.

Tab.~\ref{tab:voiding_temp_pred_results} summarizes the predictive accuracy achieved for each sensor grid configuration. The results follow a similar trend as observed for solder area fraction prediction: a single reference point yields high errors, while increasing the grid density to $3\times3$ significantly reduces both the mean absolute error (MAE) and the maximum error. However, further increasing the grid size to $4\times4$ or $5\times5$ leads to diminishing or even negative returns, with error metrics worsening due to model overfitting and poor generalization to out-of-distribution void patterns in the test set.
As in the solder area fraction prediction, using only the three standard reference points (i.e., a sparse sensor configuration) as input yields very poor results.

\begin{figure}[t]
    \centering
    \includegraphics[width=\linewidth]{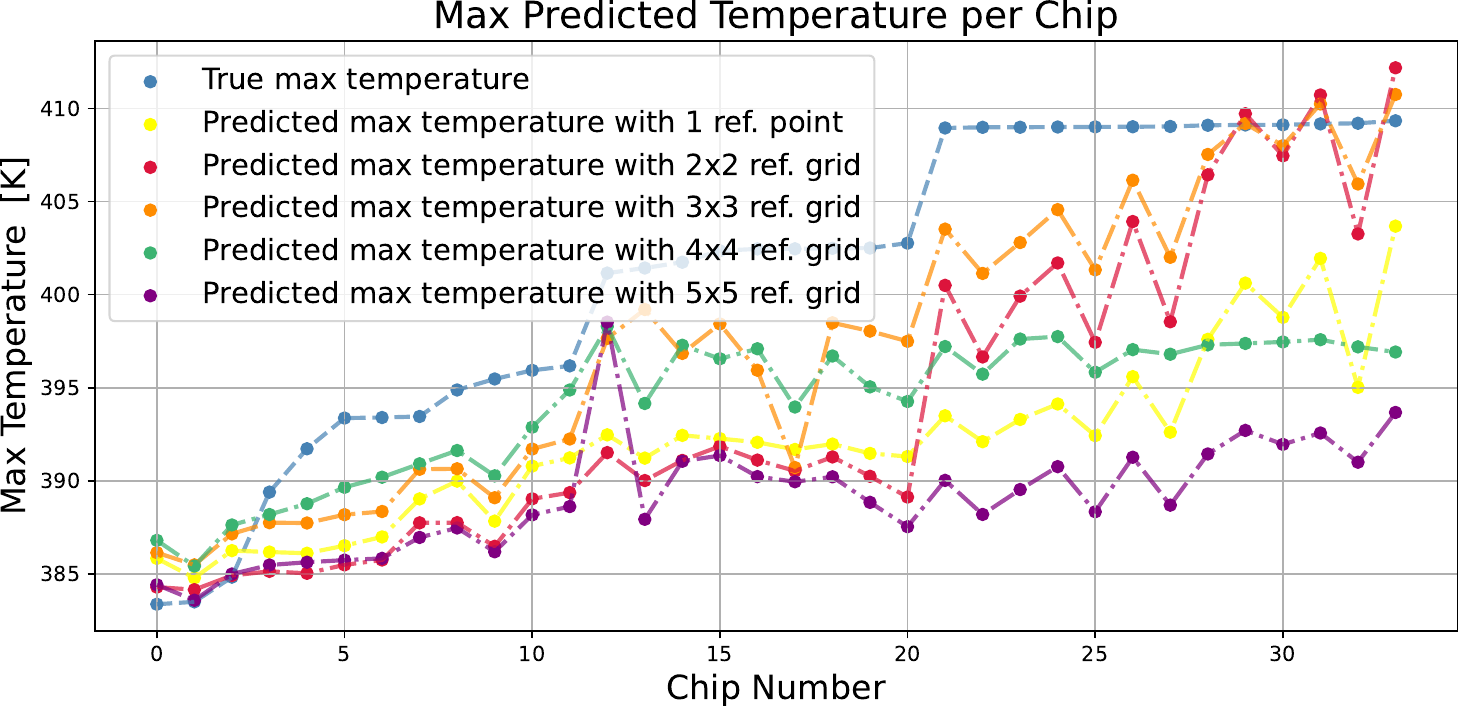}
    \caption{
        Predicted and true maximum chip temperature for different sensor grid configurations in the solder voiding scenario test set. 
    }
    \label{fig:max_temp_voids_grid}
\end{figure}

Fig.~\ref{fig:max_temp_voids_grid} visualizes the predicted and true maximum chip temperatures for all test samples and sensor configurations. As expected, the most accurate tracking of the true temperature trajectory is achieved for the $3\times3$ grid, while sparser or excessively dense grids show increased scatter and deviation, particularly in the most degraded cases.

\begin{table}[tbp]
\caption{Test set performance for maximum chip temperature prediction under voiding, as a function of sensor grid size.}
% \begin{center}
\centering
\begin{tabular}{|l|c|c|c|c|}
\hline
\textbf{Sensor grid} 
& \makecell{\textbf{MAE} \\ \textbf{[K]}} 
& \makecell{\textbf{Max Error} \\ \textbf{[K]}}
& \makecell{\textbf{Mean Rel.} \\ \textbf{L1 Error}}
& \makecell{\textbf{Max Rel.} \\ \textbf{L1 Error}} \\
\hline
$1 \times 1$  & 9.3246 & 17.2359 & 0.0231 & 0.0421 \\
$2 \times 2$  & 7.2302 & 13.6483 & 0.0180 & 0.0339 \\
$3 \times 3$  & \textbf{4.1525} & \textbf{11.7532} & \textbf{0.0103} & \textbf{0.0292} \\
$4 \times 4$  & 11.2984 & 20.8141 & 0.0279 & 0.0509 \\
$5 \times 5$  & 11.9166 & 20.8929 & 0.0294 & 0.0511 \\
\hline
$3$ reference  & 13.2575 & 21.0450 & 0.0328 & 0.0515 \\
\hline
\end{tabular}
\label{tab:voiding_temp_pred_results}
% \end{center}
\end{table}

\section{Discussion}

The present study systematically investigated the potential of machine learning–based solutions for reconstructing internal solder degradation states and temperature distributions in IGBT modules, leveraging synthetic datasets derived from finite element simulations under two canonical degradation scenarios: solder layer delamination and solder voiding. The adopted methodology followed a sequential approach, where the solder area fraction was first inferred from accessible sensor measurements and known device operating parameters, and the estimated degradation state was subsequently used as an input for temperature field prediction. This workflow not only mirrors the structure of practical condition monitoring pipelines—where the accurate assessment of device health precedes or informs further reliability analyses—but also reflects the importance of distinguishing between degradation estimation and thermal prognostics in predictive maintenance strategies for power modules.

At the heart of this approach is the recognition that monitoring the maximum chip temperature is critical for reliable prognostic and health management. Peak temperature is a sensitive indicator of localized stress, potential thermal runaway, and is directly linked to remaining useful life estimates~\cite{Hu2019, Wang2022}. In safety-critical applications, robust estimation of maximum temperature under degraded conditions is therefore central to both predictive maintenance and fault avoidance.

From a methodological perspective, the separation of solder area fraction prediction and temperature field reconstruction offers several advantages. 
It enables targeted optimization of each model stage for its specific task and input characteristics. For instance, the solder area fraction prediction benefits from compact input representations and can leverage simple sensor configurations in well-observable degradation regimes (e.g., delamination). Conversely, temperature prediction models—especially those based on GNNs—exploit spatially distributed input and domain topology to capture the complex interplay between heat flow and device geometry. By decoupling these tasks, the pipeline facilitates modular upgrades and 
detailed
diagnostics, where uncertainty at each stage can be quantified and propagated. While not explicitly addressed in the present work, this aspect will be the subject of future investigation, including the use of the devised physics-informed regularization alongside classical probabilistic models to improve the reliability of uncertainty estimates.

A notable outcome of this investigation is the clear evidence that the observability of solder degradation is fundamentally tied to the spatial configuration of available sensors. In the delamination case, where degradation proceeds in a spatially coherent and relatively predictable manner, both classical regressors and neural models perform well with a minimal input set: just three reference temperatures (including the estimated junction temperature, $T_j$) and the maximum heat source $h$. The inclusion of $h$ as an input is justified by its direct relationship to device operating conditions—power dissipation is typically either measured or can be reliably estimated in real systems through electrical monitoring, making it a physically meaningful and practically accessible input. This validates the feasibility of deploying such models in the field, as the required inputs are within reach of standard sensor setups and basic system monitoring.

In contrast, the solder voiding scenario demonstrates the limitations of sparse sensing. Here, spatially heterogeneous void formation induces local thermal anomalies 
only partially visible to widely spaced sensors. 
The sensor placement analysis clearly demonstrates a threshold effect: only when the sensor network achieves moderate coverage of the chip surface (e.g., a $2\times2$ or $3\times3$ grid at the DCB copper layer) does the prediction accuracy improve to practical levels, with larger grids offering diminishing or even negative returns due to model overfitting. This 
is likely related to the intrinsic ill-posedness of inferring distributed degradation from local measurements, as well as the curse of dimensionality in high-capacity models. These observations carry direct implications 
for 
practical monitoring systems
design. 
While minimal sensor configurations suffice for global, homogeneous degradation modes (such as delamination), spatially distributed faults like voiding necessitate more comprehensive sensor networks—possibly requiring the integration of thin-film or foil-type temperature sensors at the DCB copper layer. The feasibility and cost of such integration, as well as the long-term stability and calibration of embedded sensors, represent important engineering trade-offs for future research and industrial implementation.

A further dimension of this study is the exploration of hybrid, physics-informed models, in which domain knowledge is encoded via a regularization term derived from the heat equation. The results indicate that the inclusion of this term confers greater robustness and accuracy, especially for highly degraded cases where purely data-driven models may struggle to generalize. The physics-based regularization effectively acts as an inductive bias, promoting physically consistent predictions and mitigating the risk of overfitting to simulation artifacts or limited training data. 
Notably, the incorporation of the physics-based regularization term has the potential to reduce model complexity and data requirements by enabling smaller, more data-efficient architectures to achieve physically consistent predictions. In principle, this hybrid approach may also facilitate faster convergence during training and improved generalization from limited data. Future work should more systematically evaluate these tradeoffs, including the impact on model size, training time, and interpretability, as well as the practical constraints for real-time or embedded deployment.

Regarding the practical feasibility of the proposed solutions, several considerations emerge. The two-stage ML pipeline is well-suited for implementation on embedded platforms: the solder area fraction predictor is extremely lightweight, with average inference times per chip of $0.053\,\mathrm{ms}$ and a RAM footprint of only $0.53\,\mathrm{MB}$ (inputs are just four floating-point values), enabling real-time operation even on low-cost microcontrollers. The graph-based temperature prediction model, while more demanding (inference time per graph $\sim0.022\,\mathrm{ms}$, RAM $\sim2.4\,\mathrm{MB}$), still remains within the capabilities of modern edge processors,
especially when optimization techniques such as quantization, pruning, or knowledge distillation are applied.
The overall pipeline is thus not only accurate, but also amenable to integration in on-device health monitoring systems, with further optimization possible via model compression or hardware-specific acceleration.
It is worth emphasizing that while model training 
does require substantial computational resources (GPUs or high-performance CPUs), this phase can be performed offline using large synthetic or experimental datasets. Once trained, model weights can be deployed to control systems for efficient inference.

\section{Conclusion}  

The combination of ML-based virtual sensing, physics-informed learning, and integrated sensor networks holds substantial promise for advancing prognostic and predictive maintenance strategies in power electronics. This work presents a two-stage machine learning approach for virtual sensing in IGBT modules, enabling the estimation of solder layer degradation and internal temperature fields from a small set of physically accessible measurements. Using synthetic datasets from finite element simulations, 
we demonstrate high accuracy for both
predictive tasks.
tasks. For classical delamination, reliable results are obtained even with minimal sensor setups, whereas accurate monitoring under solder voiding requires more spatially distributed sensing. The study further demonstrates that integrating physics-informed loss functions based on the heat equation improves prediction accuracy.
Apart from what already discussed, future work will address experimental validation in real hardware, extension to multi-chip modules, additional degradation mechanisms (e.g., wire bond lift-off, metallization fatigue), assessment of long-term robustness in operational environments,  and systematic reduction of training data requirements as well as performance under realistic field conditions, including sensor noise and calibration drift.

\appendix\label{appendix_heateq}

\begin{figure}[t]
    \centering
    \includegraphics[width=0.6\linewidth]{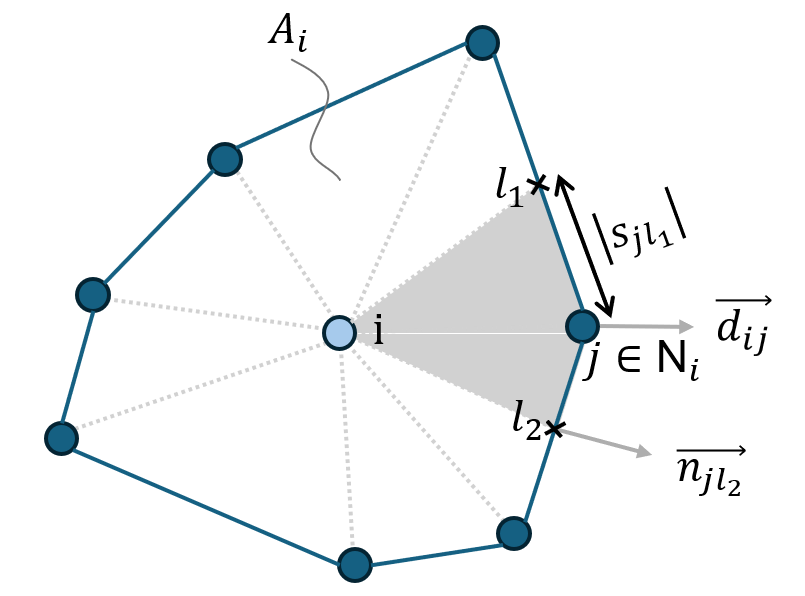}
    \caption{
        Illustration of the discretization element of node $i$ used for the definition of the physics-based loss term.
    }
    \label{fig:physics_loss}
\end{figure}

The starting point for the derivation of the physics-based loss is the differential form of the steady-state heat equation (with the heat conductivity $k$, density $\rho$, and volumetric heat source $h$) given by
\begin{equation}
    \mathcal{L}(T) = \nabla \cdot (k \vec{\nabla} T) + \rho h.
\end{equation}
The loss is defined to minimize the difference between the NN and FEM contribution of $\mathcal{L}$,
\begin{equation}
    \left|\mathcal{L}(T\lowr{NN}) - \mathcal{L}(T\lowr{FEM})\right| \rightarrow 0.
\end{equation}
The 2D graph considers only points on the top layer of the chip, thus, $k$ and $\rho h$ are the same for all points, and the corresponding terms can be neglected because the loss will be used for minimization or cancel, respectively. Thus, only the 2D-Laplace of the temperature field $\Delta T$ needs to be considered.
Next, 

First, we change to an integral formulation of the equation (line 1 in Eq.~\ref{heateq_2}), rewrite it using Gauss's theorem of integration (line 2), and discretize it for the 2D graph structure (line 3), see Fig.~\ref{fig:physics_loss}: 
for each node $i$ of the graph we construct an element surrounding it which includes all the nodes $j \in \mathcal{N}_i$ connected to it (denoted as the neighbours $\mathcal{N}_i$ of $i$). 
Note, that in contrast to a FE mesh, we obtain overlapping elements, however, contributions from overlapping volumes cancel. Thus, we only have to consider the reduced heat equation given by

\begin{align}\label{heateq_2}
 \mathcal{L}\lowr{reduced}(T) 
    & = \int_{A} \Delta_{2D} (T) dA \\
    &= \int_{\partial A}\left[ \vec{\nabla}_{2D} (T) \cdot \vec{n}\right] ds\\
    &\approx \sum_{i = 1}^N \int_{\partial A_i} \left[\vec{\nabla}_{2D} (T_i) \cdot \vec{n}_i\right] ds_i. %ds_i \delta z.
\end{align}
$\partial A$ denotes the surface of the integration volume $A$.
After discretization we obtain a sum over all nodes $N$ identified via their central node $i$ (line 3), where $\vec{n}_i$ is the surface normal of the discrete element $A_i$ which is spanned by all neighbouring nodes $j\in \mathcal{N}_i$.

For evaluation of the integral of the $i$'th element, the surface of element $A_i$ in the 2D plane is split into discrete surface segments $\Delta S_k$ comprising the circumference of the surface element. Thus, the integral can be rewritten as 
\begin{align}
   &\sum_{\Delta S_k \in \partial A_i} \left[\vec{\nabla} T \cdot \vec{n}\right]_{i} \Delta S_k   = \\ &\sum_{\Delta S_k \in \partial A_i} \begin{cases}
            \text{For internal $\Delta S_k$}: \\
             \sum_{j\in \mathcal{N}_i} (T_j - T_i) \left[ \sum_{l=1,2} \frac{\vec{d}_{ij} \cdot \vec{n}_{jl}}{|\vec{d}_{ij}|^2}  \frac{|S_{jl}|}{2}  \right] \\  \\
            \text{For $\Delta S_k$ on boundary:}\\ 
            (T\lowr{ext} - T_i) \frac{\alpha}{k} \Delta S_k,
      \end{cases}
\end{align}
where we distinguish between surface segments $\Delta S_k$ within the system, and boundary segments where the discretized version of the Robin boundary conditions is applied (see description of the FEM simulations in Sec.~\ref{sec:dataset}). For the approximation of the gradient, $\Delta S_k$ is split into two path segments $S_{jl}$ per $k$ from the node $j$ to the midpoints of the two joint adjacent nodes ($l=1$ and $l=2$) of $i$ and $j$. $\vec{d}_{ij}/|\vec{d}_{ij}|$ and $\vec{n}_{jl}$ are the (normalized) direction from node $i$ to node $j$, and the normal vector of the discretized surface element, respectively. $\alpha$ is the heat transfer coefficient.

\bibliographystyle{IEEEtran}
\bibliography{literature}

\end{document}